\documentclass[11pt]{article}

\usepackage[margin=1.2in]{geometry}
\usepackage{apacite}
\usepackage{xcolor}
\usepackage{graphicx}
\usepackage{tikz}
\usepackage{natbib}

\usepackage{lscape}

\usepackage{texshade}

\usepackage{tikz}
\usetikzlibrary{trees, positioning}

\title{An Approach for Molecular and Biological Characterizations of Virulent Influenza A Viruses}
\author{Meitner Cadena and Alejandro Yerovi}

\begin{document}

\maketitle

\begin{abstract}
We propose an approach based on a combination of physical, chemical, and mathematical methods to identify and characterize virulent influenza A viruses (IAVs) through the analysis of the hemagglutinin protein.
These methods include the isoelectric point, extreme value theory, and tree-like classification. The characterization process involves molecular and biological aspects. This procedure was applied to an IAV sample that included strains related to known influenza pandemics. The results provided clear position and amino acid pairs that identify these virulent viruses. These results show that our approach is promising to contribute new methodologies to identify and characterize virulent IAVs.
\end{abstract}

\textbf{Keywords:} influenza A virus, hemagglutinin, pH, isoelectric point, extreme value theory, tree-like classification

\section{Introduction}

It is well-known that the pH influences virus activity, when mainly this parameter decreases.
These interactions should be quite precise, since too low a pH can produce a temporal irreversible conformational change in the influenza A virus \citep{stegmann1987effects}.
Still, this modification can become permanent if exposure to low pH was for prolonged periods, leaving the virus inactive thereafter.
Viral proteins such as hemagglutinin (HA) and matrix protein 1 undergo structural changes due to low pH to facilitate the viral infection process.
This form of activation is usually achieved when the virus reaches the endosomes inside the host cells \citep{moreirainfluenza}.
Precisely, the acidic character of these large protein complexes contributes to said activation of the virus.

In normal environments, the virus cannot maintain a low pH because it risks becoming unstable.
However, some forms of this virus have shown to hold lower pH levels than usual in normal conditions.
This ability would confer to these variants more stability in acidic environments as those of endosomes, which has been linked to increase their infectivity, pathogenicity, and transmissibility.
In effect, in vitro experiments have shown that HA that has pH stability provides the virus with more efficiency for its replication \citep{singanayagam2019influenza}. We took advantage of this observation to formulate a systematic quantification of the pH of HA proteins using isoelectric point analysis, extreme value theory, and a novel hierarchical classification that guides the linkage of amino acid traits to virulence.

In practice, several procedures have been implemented to determine the virulence of influenza A viruses. In this way, protection mechanisms would be established aimed at at least avoiding variants of these viruses that could lead to serious situations such as pandemics. These procedures are based mainly on laboratory tests and assays that attempt to detect variants that present a high risk of virulence, as the molecular configurations necessary to produce these risk levels are unknown \citep{tscherne2011virulence}. 
For example, it involves identifying variants that produce changes that alter the entry of viruses as well as their binding to host cells \citep{liang2023pathogenicity}, changes in the activity of the virus polymerase \citep{tscherne2011virulence} and mediations to reduce the host's immune response \citep{kobasa2004enhanced}.
However, these laboratory studies are limited due to the availability of resources and execution time. That is, these analyzes are reduced to a few global assessments.

In this article, we propose a different approach to assess virulence. This procedure is practical since only HA sequences are used. Additionally, a molecular characterization of high-risk virulence is provided by analyzing historically observed patterns of influenza A pandemics. It is based on the identification of extreme pH values of HA proteins that would be related to high virulence, where said pH values are obtained through the notion of isoelectric point. In this way, molecular characterizations of HA proteins having extreme pH values can be obtained using classification-like trees.

The following section describes the data and analysis procedures used to identify virulent influenza A viruses and characterize them molecularly and biologically. The results of the application of the proposed analysis method are presented in Section \ref{results}. Section \ref{discussion} then discusses these results with respect to the molecular and biological characteristics of virulent influenza A viruses. The final section present concluding comments on this research.

\section{Materials and Methods}

\subsection{Data}

All hemagglutinin (HA) proteins of influenza A viruses (IAVs) that infected humans and whose sequences were reported to the National Center for Biotechnology Information (NCBI) until July 31, 2024 were considered.
The main possible subtypes of influenza A H$n$N$m$ that affect humans were included, say H1N1, H2N2, H3N2 and H5N1 \citep{mifsud2022infection}.
From these HA proteins, the ones that had a partial coding sequence, more than one stop codon or without their year of registration were discarded.
Additionally, only sequences that had the most common nucleotide length (566) were selected.
This gave 30,314 HA proteins.

\subsection{Analysis Techniques}

\subsubsection{Isoelectric Point}

To consider the pH of each HA protein, the isoelectric point (pI) of a molecule was used. This kind of point represents the pH of this molecule when it is electrically neutral \citep{Cleaves2011}. This notion has key applications as purification of proteins through their separation \citep{kosmulski2016isoelectric}, prediction of protein behavior through protein features as solubility, stability, and activity \citep{pergande2017isoelectric}, removal of viruses from monoclonal antibodies through anion exchange chromatography \citep{leisi2021impact}, and protein extraction through isoelectric point precipitation \citep{yao2023investigation}. The HA proteins were then assigned their isoelectric point calculated from their amino acid sequences using the \texttt{computePI} function of the \texttt{seqinr} \citep{seqinr} library of the R programming environment \citep{LanguageR}. This function calculates the theoretical pI based on the pK values of the amino acids reported in \citep{bjellqvist1993focusing,bjellqvist1994reference}. More precisely, each amino acid has ionizable carboxylate and ammonium groups corresponding to the carboxyl and amino groups, respectively. This produces the following acidity and basicity concentration relationships:

$$
K_a=\frac{[\textrm{H}^+]\,[\textrm{COO}^-]}{[\textrm{COOH}]}
\textrm{\quad and\quad}
K_b=\frac{[\textrm{OH}^-]\,[\textrm{NH}_3^+]}{[\textrm{NH}_2]}\textrm{,}
$$

where $[\textrm{X}]$ indicates the concentration of $\textrm{X}$, COOH is the carboxylic acid or carboxyl group and $\textrm{NH}_2$ is the amino group.
These concentrations show the dissociation levels of the molecules considered. For instance, if COOH tends to dissociate easily, the values $[\textrm{H}^+]$ and $[\textrm{COO}^-]$ will increase while $[\textrm{COOH}]$ will decrease.
Then, pI is defined by

\begin{equation}\label{pHf}
\textrm{pI}=\frac{pK_a+pK_b}{2}\textrm{,}
\end{equation}

where $pK_a=-\log_{10}(K_a)$ and $pK_b=-\log_{10}(K_b)$.

This pI calculation has limitations. For example, post-translational modifications that could alter the protein being analyzed, such as phosphorylation of glycosylation, are not considered \citep{kozlowski2021ipc}, and possible interactions between residues are not observed, as they occur during a folding process \citep{harms2009pka}. 

\subsubsection{Thresholds For Extreme Values}

A formidable and convenient fact for humans is that influenza A pandemics that have occurred are very rare. Mathematically, this means that the events of this type of pandemic are susceptible to being analyzed using the Theory of Extreme Values \citep{haan2006extreme}. More precisely, our interest is to identify the extreme pH values of HA proteins in order to characterize them. 
In this sense, 
the mean excess function $e(x)$ of a random variable $X$ can be used to estimate from where the value $x$ said function behaves as a constant when it is assumed that $X$ follows a generalized Pareto distribution $GPD(\xi,\beta)$, $0<\xi<1$ and $\beta>0$, for large values $x$.
For this distribution, this function defined by $e(x)=E(X-x|X>x)$, where $E(X)$ means the expected value of $X$ or its mean, is, say e.g. \citep{ghosh2010discussion},

\begin{equation}\label{mef}
e(x)=\frac{\beta+\xi x}{1-\xi}\textrm{.}
\end{equation}

Therefore, $e(x)$ showing linearly increasing behaviors would suggest the presence of extreme values. Then, the threshold from which this type of values would be presented would be the smallest $x$ value for which these unusual behaviors are evident. To determine the threshold for extreme pH values, the \texttt{mrlplotx} function of the \texttt{evmix} package \citep{JSSv084i05} of the R programming language was used. In addition, since the objective extreme values of pH are those that are low, the transformation $\exp(-\textrm {pH} \times \ln(10)) \times 1,000,000$ to express these extreme values as high values was applied.



\subsection{Classification Trees}

To molecularly and biologically characterize virulent IAV variants, an approach as classification trees \citep{breiman2017classification} was applied. This non-linear, non-parametric technique identifies values of certain variables that allow us to discriminate behaviors of a target variable. 
This process uses an error function that measures how well such discrimination is achieved. Fewer errors, better classification. As a result, the original data is split based on the value and variable that produce the lowest error. This procedure is applied recursively on each subset of data obtained from the set division described above. We propose a bit different approach for targeting virulent IAVs. It consisted in to analyze sub-sequences of HA proteins in order to determine the ones related to uniquely virulent IAVs. Note that from these notable sub-sequences, their initial and final amino acids always facilitate the distinction between virulent and non-virulent AIVs, but for other amino acids in these sub-sequences their function is not delved into. Next, the positions that appear most frequently were identified, as they would be key to identifying virulent IAVs.

\begin{figure}[h!]
\centering
\begin{tikzpicture}
        \node[anchor=north west,inner sep=0pt] at (0,0){\includegraphics[scale=0.90]{"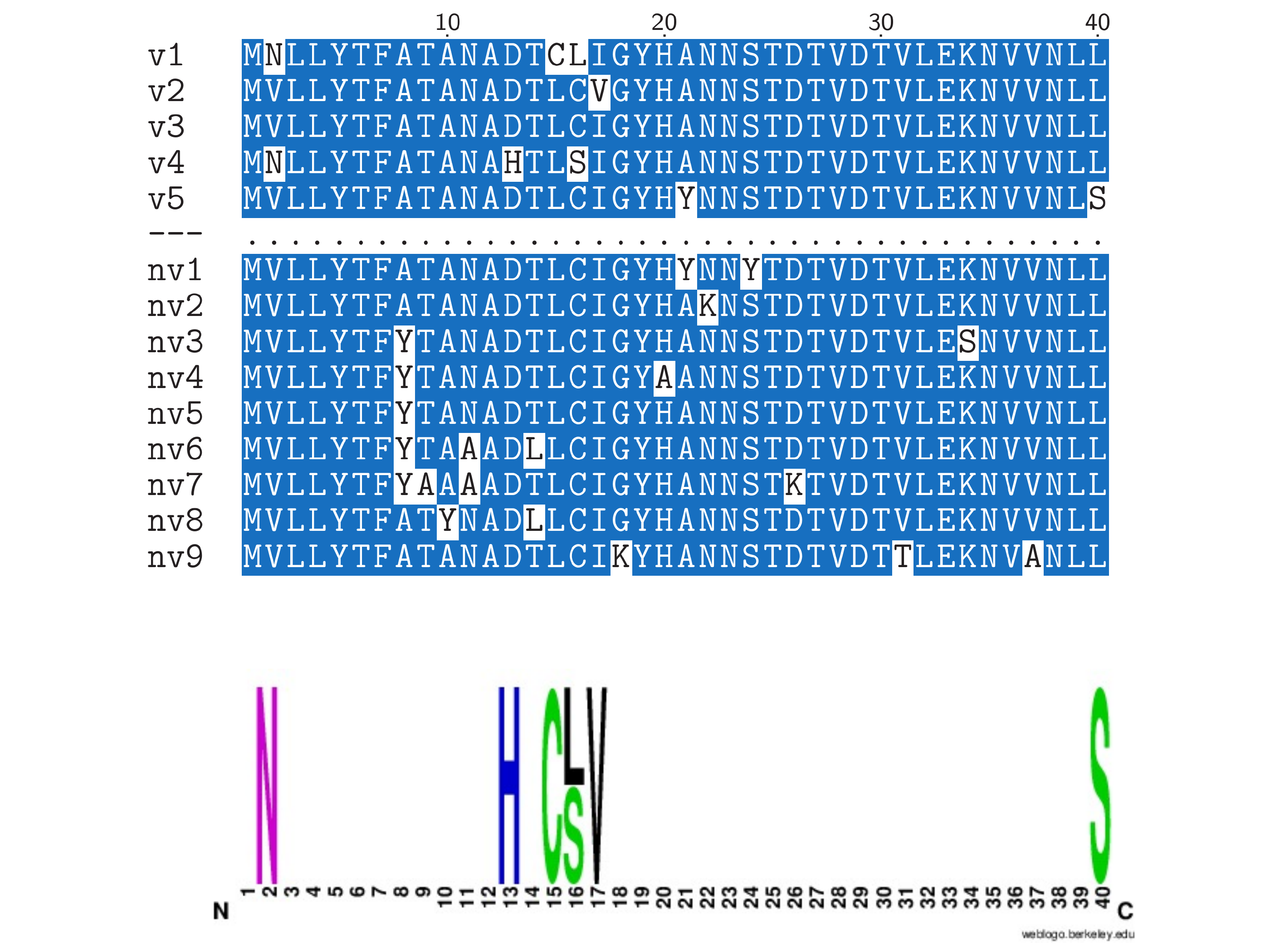"}};
        \node[font=\sffamily\bfseries\large] at (3ex,-1.5ex) {A};
        \node[font=\sffamily\bfseries\large] at (3ex,-41ex) {B};
\end{tikzpicture}
\caption{Example to identify amino acids that characterize virulent IAVs. (A) Sample of virulent and non-virulent sequences. (B) Sequence logo of potential positions and their amino acids causing virulence}
\label{example.seq}
\end{figure}

Fig. \ref{example.seq} (A) shows a sample of amino acid sequences to illustrate how to find amino acids that distinguish virulent sequences from those that do not using our classification procedure. This example shows 5 virulent, denoted by v$n$ with $n$ from 1 to 5, and 9 non-virulent, denoted by nv$n$ with $n$ from 1 to 9, HA protein sequences, which are supposed to have a length of 40. Amino acids that are not common in a given position are highlighted. Considering sub-sequences of length 1, the amino acid \texttt{N} in position 2 allows the identification of virulent AIV since no non-virulent sequence has that amino acid in position 2. Something similar happens with \texttt{Y} in position 6, \texttt{H} in position 13, \texttt{C} in position 15, \texttt{L} and \texttt{S} in position 16, \texttt{V} in position 17, \texttt{K} in position 31 and \texttt{S} in position 40. Considering sub-sequences of length 2, \texttt{CL} at position 15 is the only two-amino acid subsequence that allows the identification of a virulent AIV. These amino acids of interest are summarized in Fig. \ref{example.seq} (B) using a sequence logo \citep{dong2020overview} that was prepared using the logo tool available at \url{weblogo.berkeley.edu} \citep{weblogo}. This drawing shows the virulence-relevant amino acids and their frequencies that were identified for each position in a reference sequence.

\section{Results}
\label{results}

\begin{figure}[h!]
\centering
\includegraphics[scale=0.5]{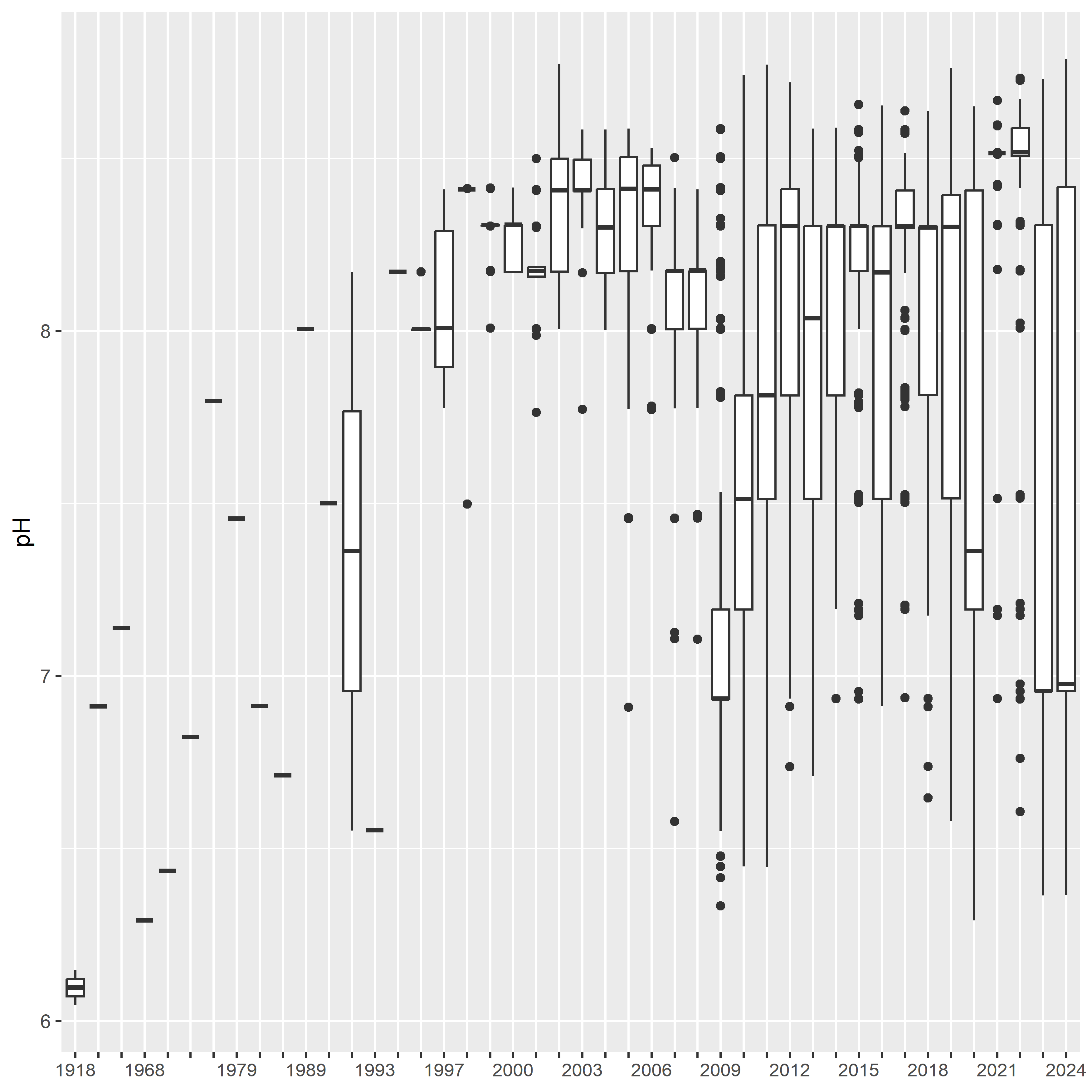}
\caption{Boxplot of analyzed pH of IAV HA from 1918 to 2024.}
\label{fig1}
\end{figure}

Fig. \ref{fig1} plots the estimated pH of the analyzed HA proteins. These values appear more frequently during the years after 2000 due to several key factors, such as advances in technology for high-throughput sequencing \citep{tsai2011influenza}, global health initiatives related to influenza \citep{petrova2018evolution}, and epidemiological studies \citep{lavenu2006detailed}. Furthermore, this figure shows that years with lower pH values would be related to IAV pandemic years, for example, the years 1918 and 2009.

\begin{figure}[htp]
\centering
\begin{tikzpicture}
        \node[anchor=north west,inner sep=0pt] at (0,0){\includegraphics[width=6cm]{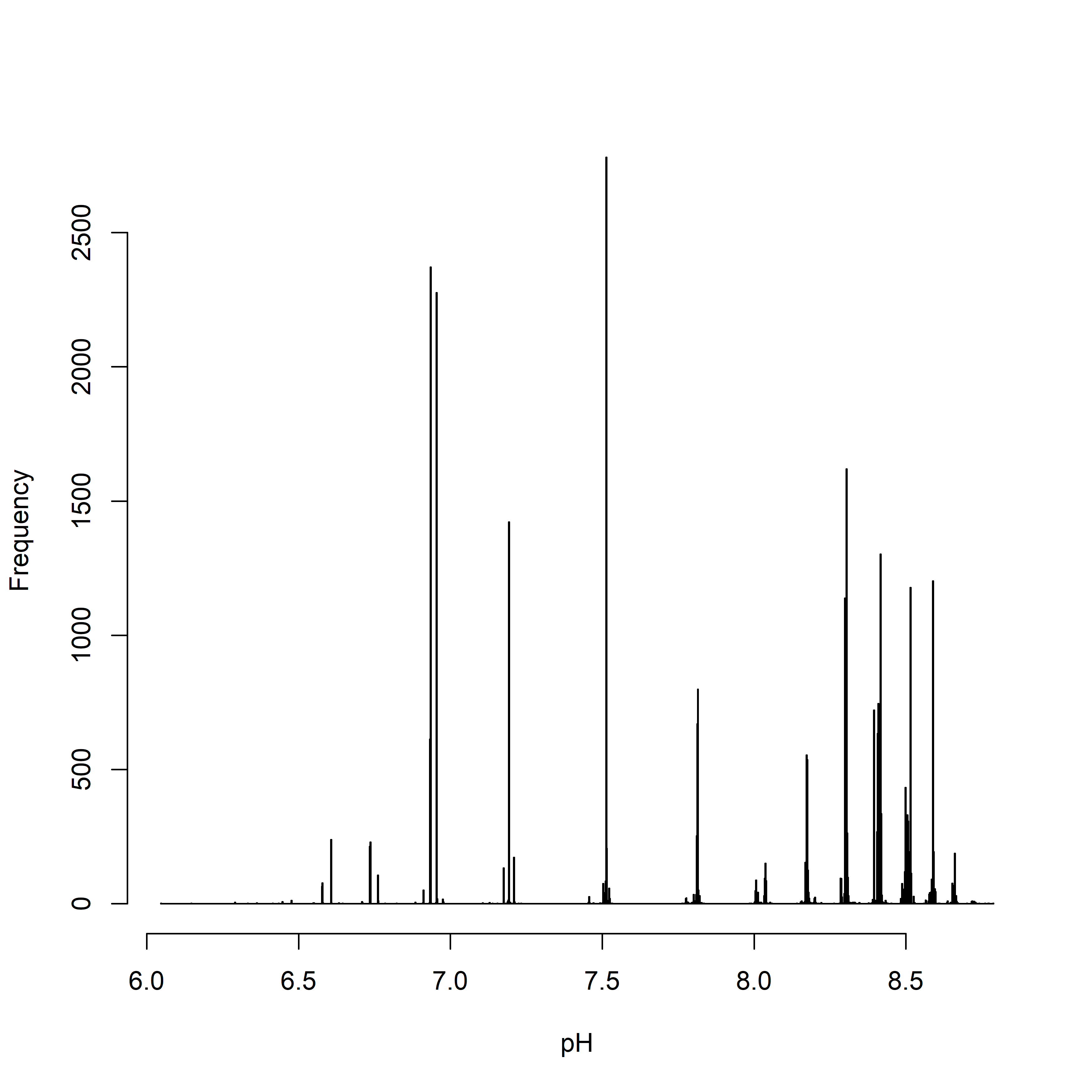}};
        \node[font=\sffamily\bfseries\large] at (-1ex,-3ex) {A};
\end{tikzpicture}
\hspace{5mm}
\begin{tikzpicture}
        \node[anchor=north west,inner sep=0pt] at (0,0){\includegraphics[width=6cm]{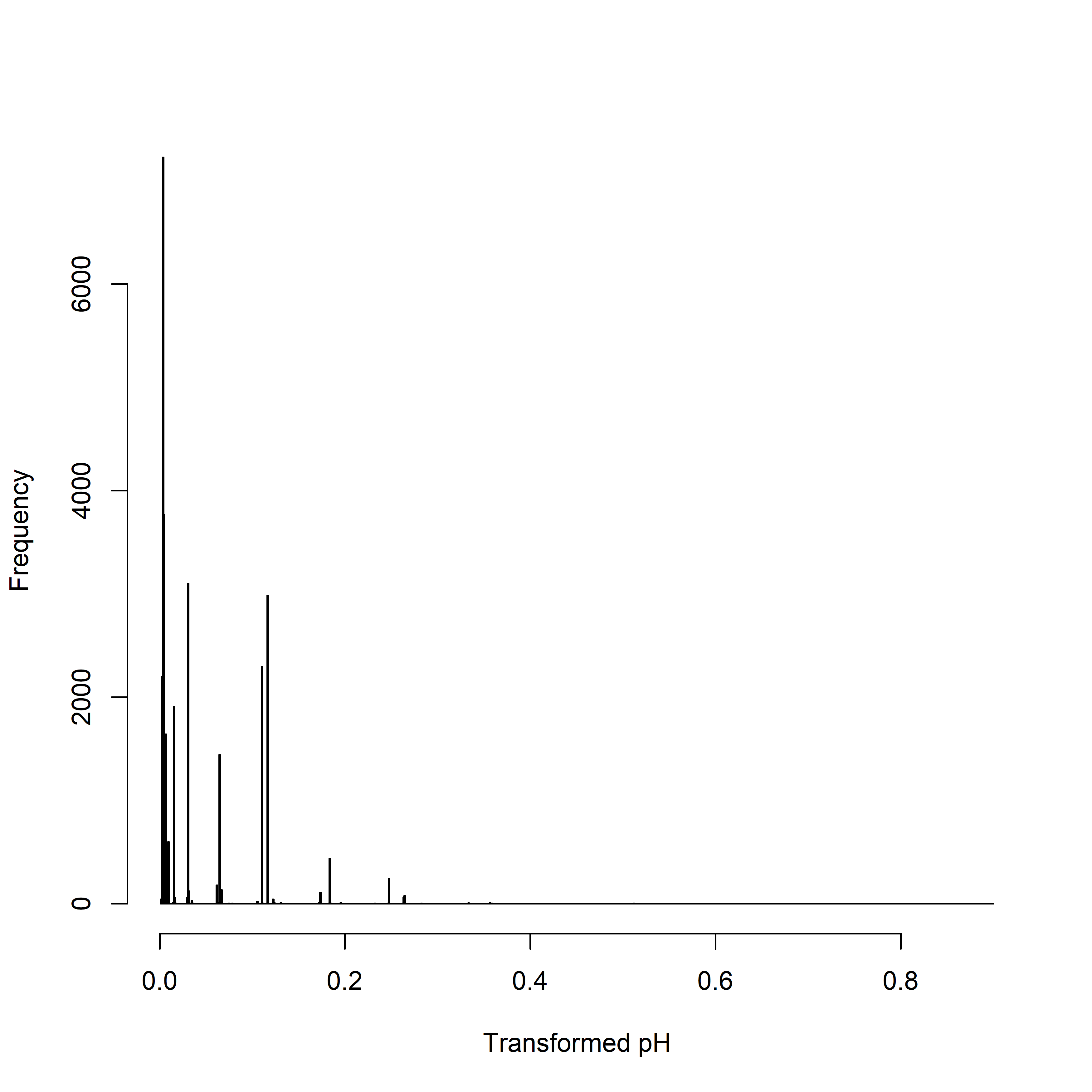}};
        \node[font=\sffamily\bfseries\large] at (-1ex,-3ex) {B};
\end{tikzpicture}
\caption{Histogram of (A) analyzed and (B) transformed pH of IAV HA from 1918 to 2024.}
\label{fig2}
\end{figure}

Fig. \ref{fig2} presents (A) all these pH values together and (B) their transformed values. The last values allow us to visualize how far the extreme values located on the right side are from the common values located on the left side. Application of the function \texttt{mrlplotx} produced the threshold 0.11639, which corresponds to pH 6.93407 (pH$^{\ast}$). Therefore, 1,666 of 30,314 HA proteins were determined to be virulent (5.5~\%).

\begin{figure}[htp]
\centering
\begin{tikzpicture}
        \node[anchor=north west,inner sep=0pt] at (0,0){\includegraphics[width=6cm]{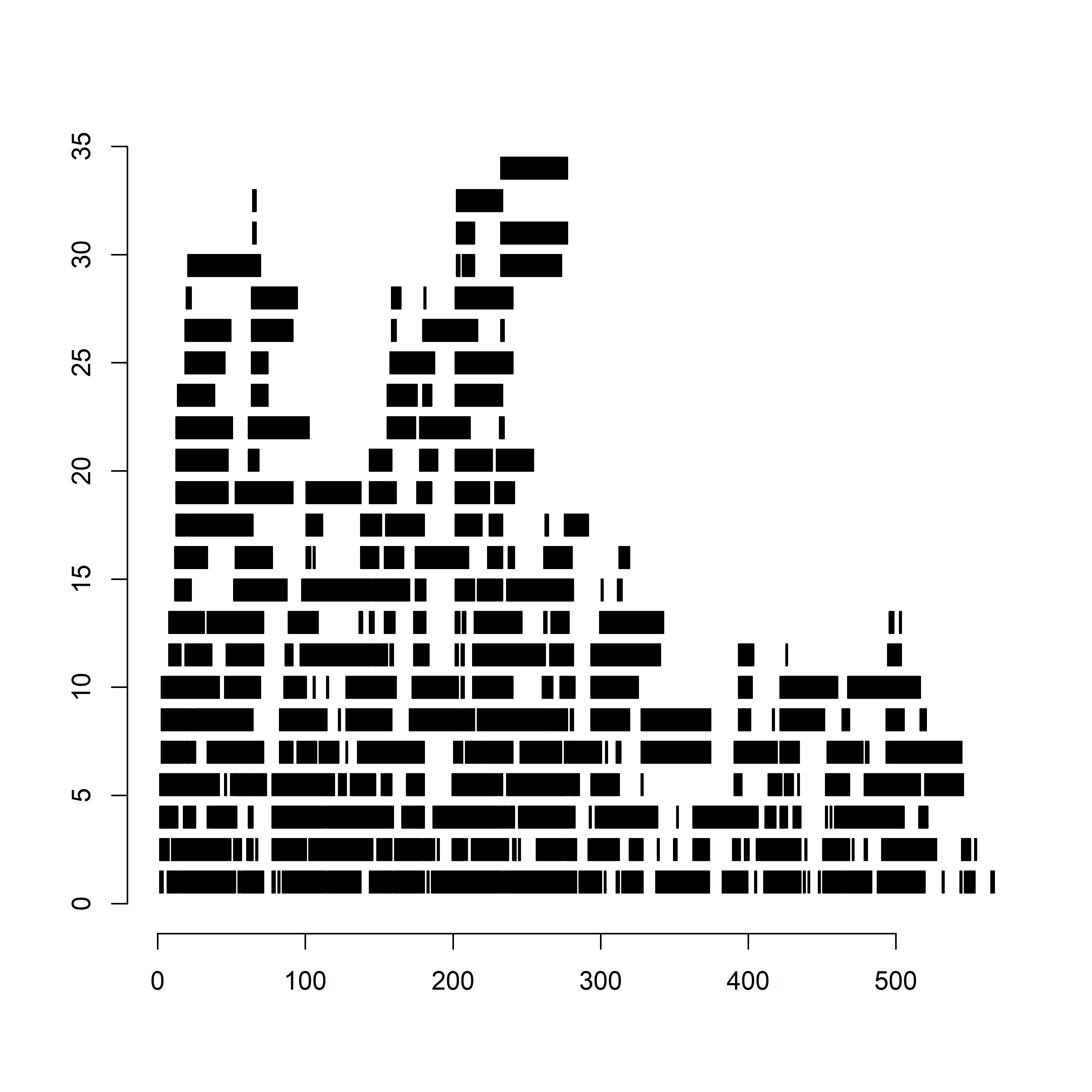}};
        \node[font=\sffamily\bfseries\large] at (-1ex,-3ex) {A};
\end{tikzpicture}
\hspace{5mm}
\begin{tikzpicture}
        \node[anchor=north west,inner sep=0pt] at (0,0){\includegraphics[width=6cm]{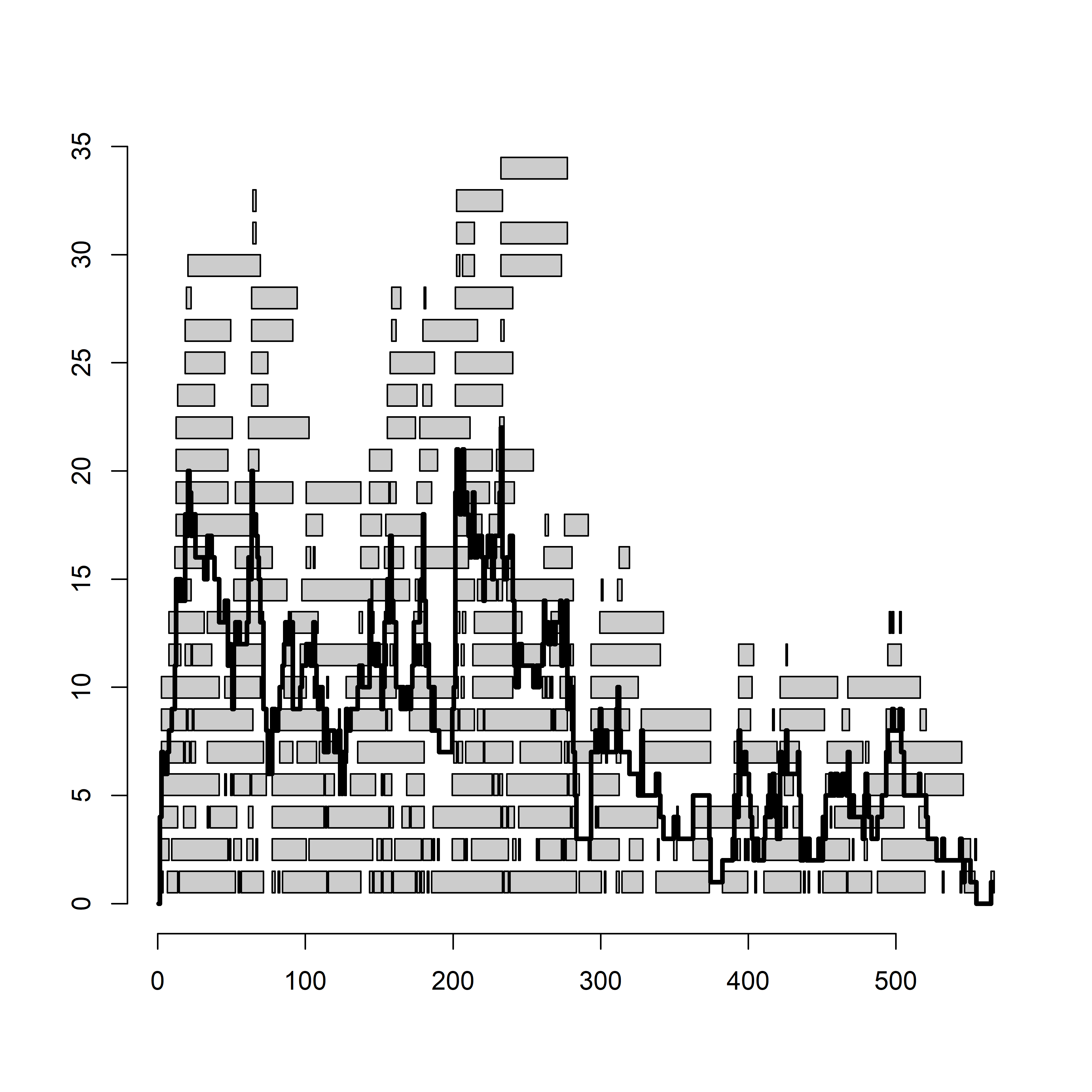}};
        \node[font=\sffamily\bfseries\large] at (-1ex,-3ex) {B};
\end{tikzpicture}
\caption{(A) Observed ranges in the sequence of HA proteins and (B) coverage (in black) of these ranges.}
\label{fig3}
\end{figure}

%
Next, the set of all sequences was analyzed using the classification approach described above to distinguish the ones with pH lower than pH$^{\ast}$ from those with higher pH. To this end, application of our procedure to classify virulent HA proteins with sub-sequence lengths from 1 to 50 allowed the identification of 284 virulent IAVs. Fig. \ref{fig3} (A) shows these results of classification, where all ranges presented in this figure correspond to those identified virulent IAVs. Fig. \ref{fig3} (B) shows the coverage for those ranges that means how many ranges overlap each position. Ranges and coverage were computed using the IRanges package \citep{pages2013package}. This coverage curve allows the visualization of positions with higher frequencies through all positions of analyzed HA proteins. This means that such positions are common to a number of virulent IAVs, which would thus be associated with key positions that have roles in virulence as they have relevant frequencies.

Table \ref{tabx1} shows the detected modes (positions) mentioned above. Their biological functionalities following e.g. \cite{sriwilaijaroen2012molecular} are indicated. The ten most frequent positions are in bold.


\begin{landscape}

\begin{table}[htp]
\footnotesize{
\centering
\begin{tabular}{lllll}
\hline
\multicolumn{1}{c}{Region} & \multicolumn{1}{c}{Domain} & \multicolumn{1}{c}{Positions}  & \multicolumn{1}{c}{Found positions} & \multicolumn{1}{c}{Biological functionalities} \\
\hline
HA1 & Fusion & 22 -- 34 & \textbf{23} & Unknown \\
 & & 34 -- ? ($<63$) & \textbf{36} & Glycosylation site found in all almost H1 viruses or unknown \\
\cline{2-5}
 & Esterasa & 64 -- ? ($<76$) & \textbf{66} & Glycosylation site found only in seasonal H1 viruses or unknown \\
 & & 84 -- 95 & 88 & Antigenic site less changeable$^{\dag}$ \\
 & & 95 -- 115 & 103 & Conserved region$^{\ddagger}$ \\
\cline{2-5}
 & Receptor-binding & 133 -- 138 & 137 & Monoclonal antigenic site$^{\dag}$ \\
 & & 143 -- 152 & 145 & Monoclonal antigenic site$^{\dag}$ \\
 & & 157 -- ? ($<163$) & \textbf{158} & Glycosylation site found only in seasonal H1 viruses or antigenic site \\
 & & 173 -- 182 & \textbf{178} & Conserved region$^{\ddagger}$ \\
 & & 198 -- 206 & \textbf{206} & Positive selected site$^{\ast}$ \\
 & & 208 -- 218 & \textbf{214} & Antigenic site$^{\circ}$ \\
 & & 228 -- 239 & \textbf{233} & Conserved region$^{\ddagger}$ \\
 & & 241 -- 263 & 250 & Conserved region$^{\ddagger}$ \\
\cline{2-5}
 & Esterase & 263 -- 276 & \textbf{264}, \textbf{272} & Glycosylation site found only in seasonal H1 viruses, positive selected site$^{\ast}$ for 264 \\
\cline{2-5}
 & Fusion & 290 -- 324 & 300, 313 & Conserved region$^{\ddagger}$ \\
\hline
HA2 & Fusion & 329 -- 483 & 337, 351, 368, 397, 416, & Conserved region$^{\ddagger}$ excepting 458 \\
 & & & 426, 443, 458, 466, 480 & \\
\cline{2-5}
 & Not defined & 489 -- 514 & 499 & Antigenic site$^{\circ}$ \\
 & & 514 -- 542 & 516, 542 & Transmembrane \\
\hline
\multicolumn{5}{l}{($\dag$) \cite{nakajima2005accumulation}, ($\ddagger$) \cite{ghafoori2023structural}, ($\ast$) \cite{ghafoori2023structural}, ($\circ$) \cite{luczo2024epitopes}}
\end{tabular}
\caption{Biological functionalities of positions of HA proteins detected for having potential roles in virulence.}
}
\label{tabx1}
\end{table}
\end{landscape}

Table \ref{tabx1} shows that the modes involved are distributed throughout the HA protein. Among their biological functions, antigenic sites stand out, suggesting that evasion of the host's immune system is an important component of the virulence strategy by lowering the pH.

\begin{figure}[h!]
\centering
\begin{tikzpicture}
        \node[anchor=north west,inner sep=0pt] at (0,0){\includegraphics[scale=0.62]{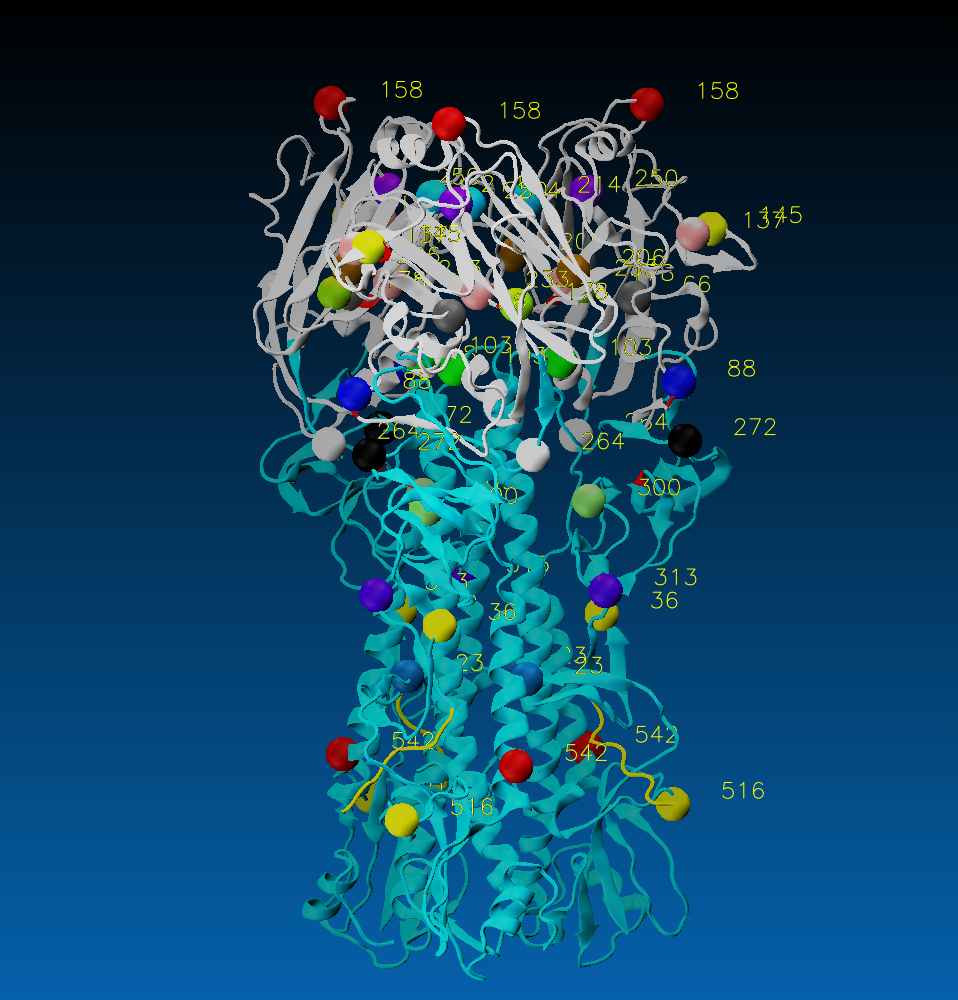}};
        \node[font=\sffamily\bfseries\large, text=white] at (1.2ex,-2ex) {A};
\end{tikzpicture}
\hspace{-2mm}
\begin{tikzpicture}
        \node[anchor=north west,inner sep=0pt] at (0,0){\includegraphics[scale=0.62]{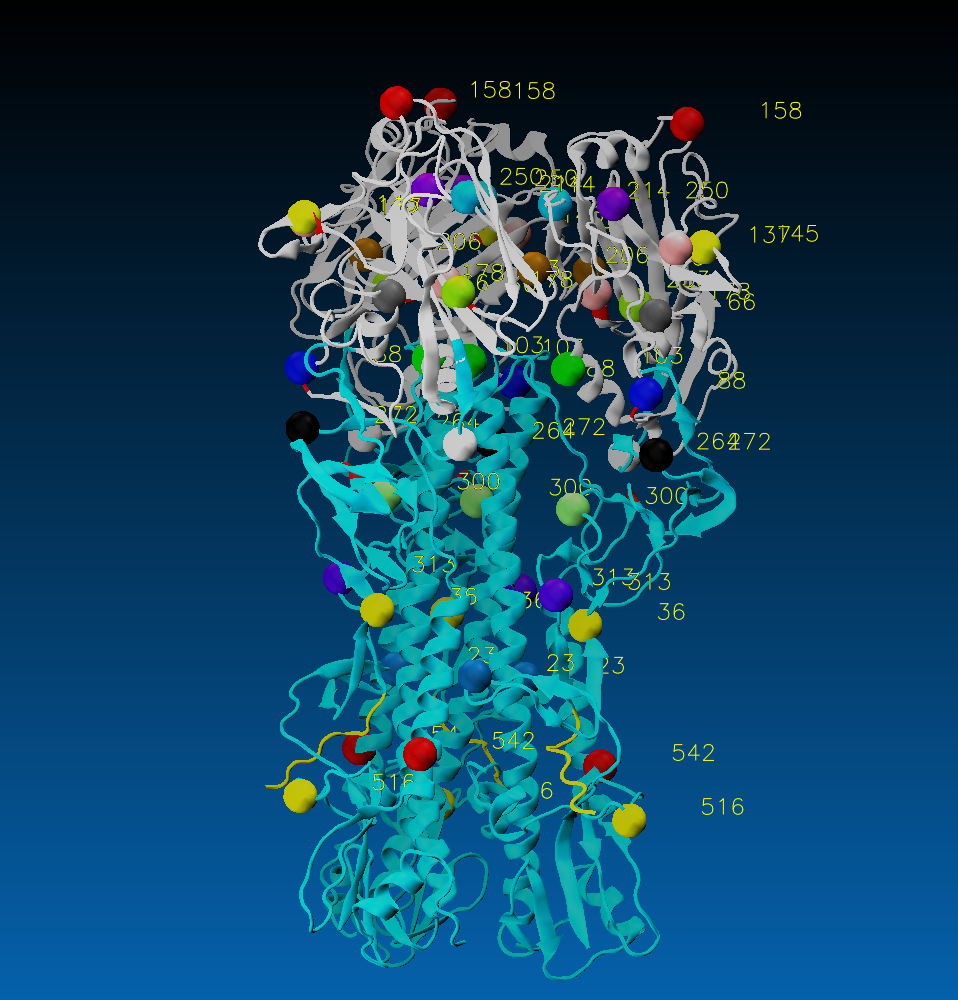}};
        \node[font=\sffamily\bfseries\large, text=white] at (1.2ex,-2ex) {B};
\end{tikzpicture}
\hspace{-2mm}
\begin{tikzpicture}
        \node[anchor=north west,inner sep=0pt] at (0,0){\includegraphics[scale=0.62]{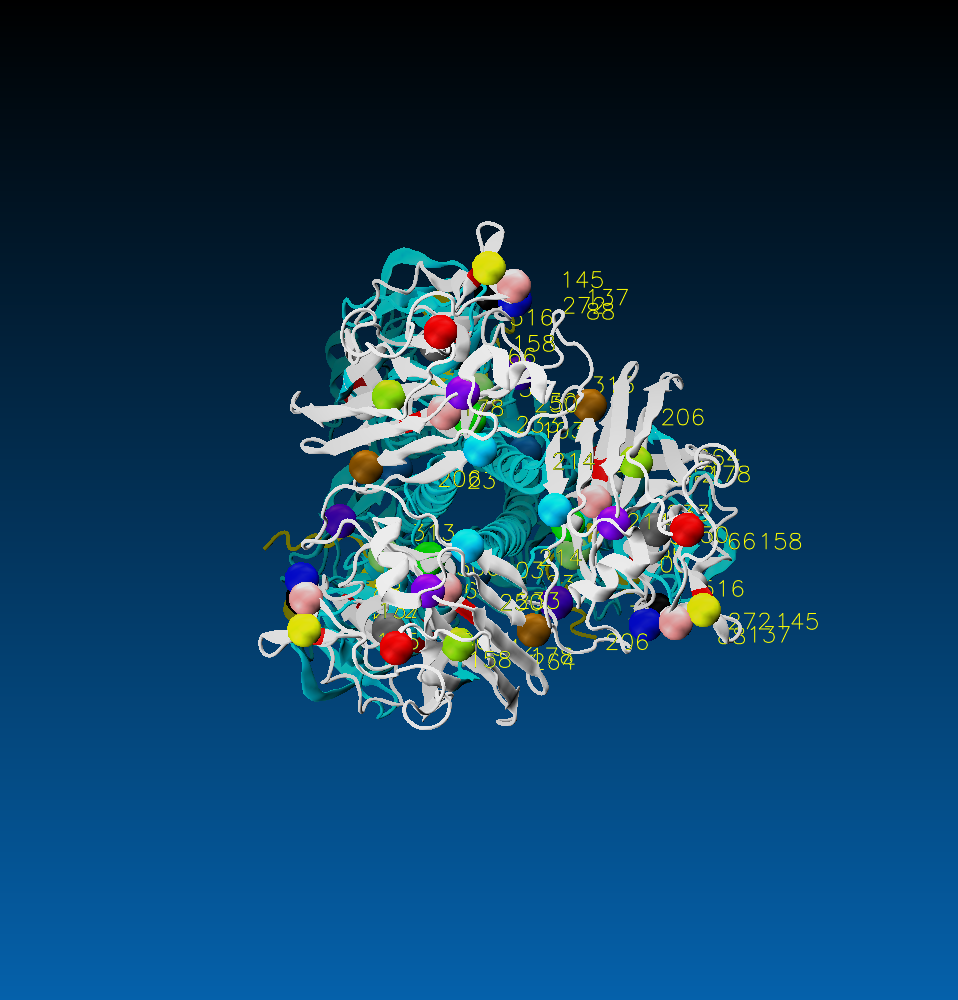}};
        \node[font=\sffamily\bfseries\large, text=white] at (1.2ex,-2ex) {C};
\end{tikzpicture}
\caption{(A) front, (B) side, and (C) top views of HA protein containing potential virulence-related positions of IAVs}
\label{img}
\end{figure}

Fig. \ref{img} shows the possible virulence-related positions that were found (in yellow) in three-dimensional representations of the HA protein. These graphs are based on the 1RUZ 1918 variant available in the Protein Data Base database \citep{wwpdb} and to obtain these graphs the VMD \citep{HUMP96} application has been used. These images show (A) front, (B) side, and (C) top views of this protein.

Regarding the amino acids associated with ten positions with highest frequencies, Table \ref{aaf} shows these results, which include the frequencies of the positions in question and also the frequencies of the amino acids involved. The positions are sorted by decreasing frequency and in ascending order by position if they match in frequency. The amino acids of each position are arranged in descending order according to their frequency and alphabetically if they coincide in frequency. These notable positions were highlighted in Table \ref{tabx1} in bold.

\begin{table}[h!]
\centering
\footnotesize{
\begin{tabular}{rrcr}
\hline
 & & \multicolumn{2}{c}{Amino acid} \\
\cline{3-4}
\multicolumn{1}{c}{Position} & \multicolumn{1}{c}{Frequency} & \multicolumn{1}{c}{Single} & \multicolumn{1}{c}{Frequency} \\
 & & \multicolumn{1}{c}{letter code} & \\
\hline
233 & 22 & I & 13 \\
 & & A & 6 \\
 & & T & 3 \\
\hline
206 & 20 & Q & 11 \\
 & & E & 4 \\
 & & D & 3 \\
 & & K & 1 \\
 & & N & 1 \\
\hline
214 & 19 & A & 13 \\
 & & S & 3 \\
 & & T & 3 \\
\hline
66 & 18 & P & 10 \\
 & & E & 6 \\
 & & K & 2 \\
\hline
23 & 17 & D & 8 \\
 & & G & 8 \\
 & & E & 1 \\
\hline
36 & 17 & V & 15 \\
 & & I & 2 \\
\hline
\end{tabular}
\hspace{2mm}
\begin{tabular}{rrcr}
\hline
 & & \multicolumn{2}{c}{Amino acid} \\
\cline{3-4}
\multicolumn{1}{c}{Position} & \multicolumn{1}{c}{Frequency} & \multicolumn{1}{c}{Single} & \multicolumn{1}{c}{Frequency} \\
 & & \multicolumn{1}{c}{letter code} & \\
\hline
158 & 17 & A & 5 \\
 & & K & 4 \\
 & & E & 2 \\
 & & G & 2 \\
 & & R & 2 \\
 & & S & 2 \\
\hline
178 & 14 & L & 7 \\
 & & P & 3 \\
 & & I & 2 \\
 & & Q & 1 \\
 & & S & 1 \\
\hline
264 & 13 & N & 7 \\
 & & T & 4 \\
 & & K & 2 \\
\hline
272 & 13 & F & 9 \\
 & & G & 4 \\
\hline
\multicolumn{4}{c}{} \\
\multicolumn{4}{c}{} \\
\multicolumn{4}{c}{}
\end{tabular}
}
\caption{Amino acids that distinguish virulent IAVs, associated to positions with highest frequencies.}
\label{aaf}
\end{table}

\section{Discussion}
\label{discussion}

Our main objective of knowing how virulent the structure of an IAV can be to plan prevention measures against possible epidemics and pandemics has been met. Indeed, our approach based mainly on the use of pH measurements, the application of extreme value theory and the configuration of classification-like trees has allowed the identification of key positions in HA proteins to describe their virulence. According to our approach, well-known pandemic years appear to be related to virulent viruses collected in 1918 and 2009, but also potential virulent IAVs would also have appeared in 1968, 1972, 2021, 2023 and 2024. Further, such identification of those key positions has been determined using local modes because the high frequency in a representative sample of virulent IAVs would mean that these positions play a crucial role. Interestingly, these positions are distributed throughout the HA protein, but some of them definitely seem very important. Furthermore, some of the specific amino acids involved in these positions have been established.

The distribution of potential virulence-related positions in Table \ref{tabx1} and the position frequency presented in Table \ref{aaf} shows that virulent AIVs would be related to both HA1 and HA2, but with more emphasis with HA1 than with HA2. This clarification between HA1 and HA2 seems not to be included in the scientific literature where what is considered is that both HA1 and HA2 would play crucial roles in the virulence of IAVs \citep{shirvani2020contributions}. On the other hand, considering the frequency of the biological functionalities of the potential positions of greatest virulence, it is observed that the increase in the intervention of the antigenic sites would be supported by the decrease in pH. This suggests that these positions would have biological functions distinct from antigenic sites before mutations occur that confer antigenic functions to these sites.

Also, it seems difficult to identify crucial positions related to virulence since this concept has not been characterized in terms of positions. In this way, our model clearly contributes to clarifying what these potential positions would be.

\section{Conclusions}

The uncertainty about how the HA protein of an IAV mutates represents a serious limitation to knowing how virulent this virus can be. Despite this, it has been possible to identify positions that would be playing a crucial role in the emergence of virulent IAVs. This analysis based on the application of physical, chemical and mathematical processes thus seems to be a promising strategy to complement the analyzes carried out to determine the virulence of an AIV.

Moreover, although the characterization of the mutations has not been addressed in this study, this essential knowledge could also be revealed by the procedure we have proposed. In fact, the identification of potential positions with specific amino acids for obtaining virulent IAVs would allow analyzing the relationships between virulent and non-virulent IAVs. Therefore, some of these relationships could involve the aforementioned mutations.

As there is still a need to better understand the composition and function of structural and antigenic integrity sites and even their evolution in positions and amino acids when virulent IAVs are formed, further investigations are required. These activities that aim at acquiring new knowledge may also include explorations on topics as Gibbs free energy and molecular symmetry among others.

\bibliographystyle{apacite}
\bibliography{references.bib}

\end{document}